\documentclass{article} % For LaTeX2e
\usepackage{iclr2024_conference,times}

\usepackage{amsmath,amsfonts,bm}

\def\eqref#1{equation~\ref{#1}}
\def\1{\bm{1}}

\DeclareMathAlphabet{\mathsfit}{\encodingdefault}{\sfdefault}{m}{sl}
\SetMathAlphabet{\mathsfit}{bold}{\encodingdefault}{\sfdefault}{bx}{n}

\usepackage{hyperref}
\usepackage{url}
\usepackage{graphicx}
\usepackage{algorithm}
\usepackage{algpseudocode}
\usepackage{listings}
\usepackage{booktabs}
\usepackage{subcaption}
\usepackage{adjustbox}

\title{A novel methodological framework for the analysis of health trajectories and survival outcomes in heart failure patients}

\author{Juliette Murris$^{1}$, Tristan Amadei$^{2}$, Tristan Kirscher$^{2}$, Antoine Klein$^{2}$, \\
\textbf{Anne-Isabelle Tropeano$^{3}$ \& Sandrine Katsahian$^{1,3}$} \\
$^{1}$ HeKA, Inserm, Inria, Université Paris Cité, Pierre Fabre R\&D $^{2}$ ENSAE, IP de Paris, \\
$^{3}$ CIC-1418, HEGP, AP-HP, Paris, France\\
\texttt{juliette.murris@inria.fr}
}

\iclrfinalcopy % Uncomment for camera-ready version, but NOT for submission.
\begin{document}

\maketitle

\begin{abstract}
Heart failure (HF) contributes to circa 200,000 annual hospitalizations in France. With the increasing age of HF patients, elucidating the specific causes of inpatient mortality became a public health problematic. We introduce a novel methodological framework designed to identify prevalent health trajectories and investigate their impact on death. The initial step involves applying sequential pattern mining to characterize patients' trajectories, followed by an unsupervised clustering algorithm based on a new metric for measuring the distance between hospitalization diagnoses. Finally, a survival analysis is conducted to assess survival outcomes. The application of this framework to HF patients from a representative sample of the French population demonstrates its methodological significance in enhancing the analysis of healthcare trajectories.
\end{abstract}

\section{Motivating example}
Heart failure (HF) is a cardiovascular condition characterized by the heart's inability to pump sufficient blood to meet the body's oxygen and nutrient needs. It is a prevalent disease,  affecting 1 to 2\% of adults in developed countries, and around 64 million people worldwide \citep{savarese2022global}. Chronic HF often goes with repeated hospitalizations and embodies the condition with highest 30-days re-hospitalization rate \citep{constantinou2021patient}. In France alone, over 1.5 million individuals suffer from HF, resulting in approximately 200,000 hospitalizations annually. Thus, understanding primary causes of death in these patients and identifying their most frequent inpatient trajectories holds significant potential for public health impact.

\begin{figure}[h]
    \centering
    \includegraphics[width=0.4\linewidth]{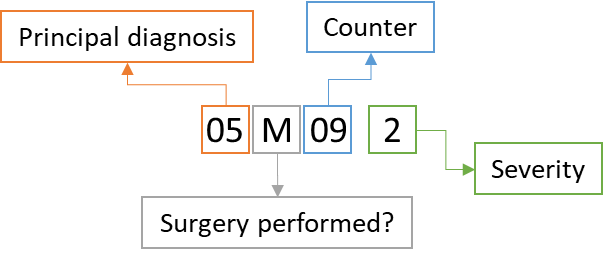}
    \caption{ICD-10 architecture}
    \label{fig:ghm}
\end{figure}

Electronic health records in France aggregate data from all hospitalizations. The EGB (\textit{Echantillon Généraliste des Bénéficiaires}) is a random sample and representative of  1/97th of the population over a two year follow-up \citep{de_roquefeuil_lechantillon_2009}. Trajectories are established using primary and associated diagnoses based on the International Classification of Disease (10th edition (ICD-10)). ICD-10 code syntax is text-based, and includes the principal diagnosis category, indication of surgical procedures, a counter, and a severity indicator (example given in Figure \ref{fig:ghm}, '05M092' is the ICD-10 code for HF).

The primary objective of this work is to identify frequent inpatient health trajectories in HF patients in France. Secondary objectives include the investigation of their associations with mortality. 

\section{Related work}
The analysis of trajectories in healthcare is a complex task which helps understanding the evolution of patients pathway over time. Many approaches deal clustering \citep{chouaid2022machine,lambert2023tracking,zhong2021clustering}, and others with times series, like Markov chains \citep{macdonald1997hidden} or neural networks \citep{hewamalage2021recurrent}. More recently the advent of text-based input enabled to collect more and more information \citep{nguyen2018effective, zhu2021embedding}.

However, none of these methodologies combine pattern mining, clustering and survival analysis all together, even though these concepts are individually well-established and active research fields \citep{kang2020prefixspan,leis2023k,murris2023towards}. This work is a demonstration of the assembling of existing tools to answer concrete clinical need and represents a "bridging fields" contribution.

\section{Methodological framework}
\subsection{Sequential pattern mining - the \textit{PrefixSpan} algorithm} \label{sec:ss-spm}
We used sequential pattern mining technique to extract frequent trajectories \citep{masseglia_extraction_2004}. \textit{PrefixSpan} algorithm uses sequence patterns as tuples $(P, sup)$, where $P$ is a sequential pattern and $sup$ is the number of sequences in the database that contain $P$ \citep{pei_prefixspan_2001}. The sequence patterns are used to efficiently compute the support of candidate patterns and avoid unnecessary database scans. The algorithm efficiently identifies frequent sequential patterns in the database while maintaining a concise representation of the patterns (see Algorithm \ref{alg:cap}).

\subsection{Unsupervised learning with patient clustering} \label{sec:ss-clustering}
The clustering of patients brings the interpretability layer to the approach \citep{pinaire_explorer_2017}. Clusters should fit closely patients' trajectories based on their successive hospitalization sequences (see Algorithm \ref{alg:kma}).  K-medoids algorithm was used to deal with string data and aims to partition input into $k$ clusters \citep{kaur2014k}. Each cluster is represented by a single data point called medoid. 

\begin{figure}[h]
    \centering
    \includegraphics[width=0.53\linewidth]{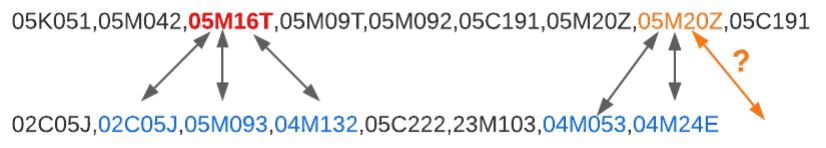}
    \caption{Measuring distances for two patient sequences}
    \label{fig:ghm_dist}
\end{figure}

Levenshtein distance $\mathcal{D}_L$ was used to measure the distance between two hospitalization sequences \citep{yujian2007normalized}. Traditionally, the Levenshtein distance calculates the minimum number of edits to perform on single-characters to transform the word $a$ into the word $b$:
\begin{equation*}
    \mathcal{D}_L(a,b) = 
    \begin{cases}
        \max (|a|,|b|) & \text{if } \min (|a|,|b|) = 0, \\
        lev(a_{1:},b_{1:}) & \text{if } a[0] = b[0], \\
        1 + \min \begin{cases}
            lev(a_{1:},b) \\
            lev(a,b_{1:}) \\
            lev(a_{1:},b_{1:})
        \end{cases} & \text{otherwise.}
    \end{cases}
\end{equation*}
with $|\cdot|$ the number of letters in the word, $\cdot_{1:}$ the word without its first letter and $\cdot[0]$ the first letter of the word. Throughout our study, we will only use the Levenshtein ratio, being the normalized value with $lev(a,b) = \frac{\mathcal{D}_L(a,b)}{\max (|a|,|b|)}$ with words of the same number of characters, hence $\max (|a|,|b|) = |a| = |b|$. We used this string-distance to implement a new metric $\mathcal{D}_{ICD10}$ to compute the distance between two ICD-10 codes $A$ and $B$:
\begin{align*}
    \mathcal{D}_{ICD10}(A,B) =  & \omega_1 \times lev(A_{0:2},B_{0:2}) \\
                                & + \omega_2 \times lev(A[2],B[2]) \\
                                & + \omega_3 \times lev(A_{3:5},B_{3:5}) \\
                                & + \omega_4 \times lev(A[5],B[5])
\end{align*}
with $\Omega = (\omega_i)_{i\in[1,4]} \in \mathbb{N^+}$. This weighted distance thus leverages the information from each ICD-10 component (as per Figure \ref{fig:ghm}). Then, we compare the $i^{th}$ ICD-10 code of a patient with the $(i-1)^{th}$, $i^{th}$ and $(i+1)^{th}$ ICD-10 codes of another patient, compute the distances and keep the minimum in order to get the distance between two patient sequences (Figure \ref{fig:ghm_dist}). 

The distance $\mathcal{D}_P$ between two patient sequences is the sum of all ICD-10 codes to one another. This distance respects the symmetry assumption and $\mathcal{D}_P(patient_i,patient_j) = \mathcal{D}_P(patient_j, patient_i)$. Further details on the distance matrix are given in Appendix.

Two hyperparameters require settings and are under constraint: $\Omega$ the weights of the distance metric, with $0 \leq \omega_4 \leq \omega_3 \leq \omega_2 \leq \omega_1 \leq 100$, and $k \in [2,20]$ the number of clusters. Cross-validation was used to find optimal hyperparameters using Optuna \citep{akiba2019optuna}. The score $S$ to be maximised was defined as follows, for each cluster $k$:
\begin{equation*}
    s_{p,k} = \sum_{p=1}^{N_p} (P,sup)[p]_k - (P,sup)[p]
\end{equation*}
And $S = \frac{1}{N_p \times N} \sum_{k=1}^N \sum_p^{N_p} s_{p,k}$. We set $N_p = 3$. Based on outputs from the \textit{PrefixSpan} algorithm, the idea is to determine the most frequently occurring ICD-10 code patterns of lengths {1, 2, 3} within each cluster, along with their respective frequencies. Subsequently, we calculate the frequency of these patterns across the entire dataset and compute the difference between the two. We then get the mean of these differences separately for patterns of length 1, 2, and 3. The clustering score $S$ is then established by averaging these means.

\subsection{Survival analysis} \label{sec:ss-surv}
For each patient, survival time and status are available. We implemented two ensemble methods, namely random survival forests (RSF) and survival gradient boosting. These methods are survival analysis counterparts of random forests and gradient boosting algorithms tailored for censored data \citep{ishwaran_random_2008,hothorn2006survival}. Cross-validation was performed for hyperparameter optimization.

Two metrics were used for evaluation. We assessed the goodness of fit using the Akaike information criterion (AIC), with lower values indicating better fit \citep{hu2007akaike}. Besides, the concordance index (C-index) is a generalization of the ROC-AUC that can effectively handle censored data, providing a reliable ranking of survival times based on individual risk scores \citep{harrell1982evaluating, harrell1984regression}. 

\section{Results}
This study included adult HF patients who experienced their initial HF hospitalization between 2010 and 2016. The cohort comprised 10,051 patients, accounting for a total of 85,594 hospitalizations. Five clusters were identified with optimal hyperparameters ($\Omega^T = [85,75,55,40]$). 

\begin{figure}[H]
\centering
\begin{subfigure}{.45\textwidth}
    \centering
    \includegraphics[width=\linewidth]{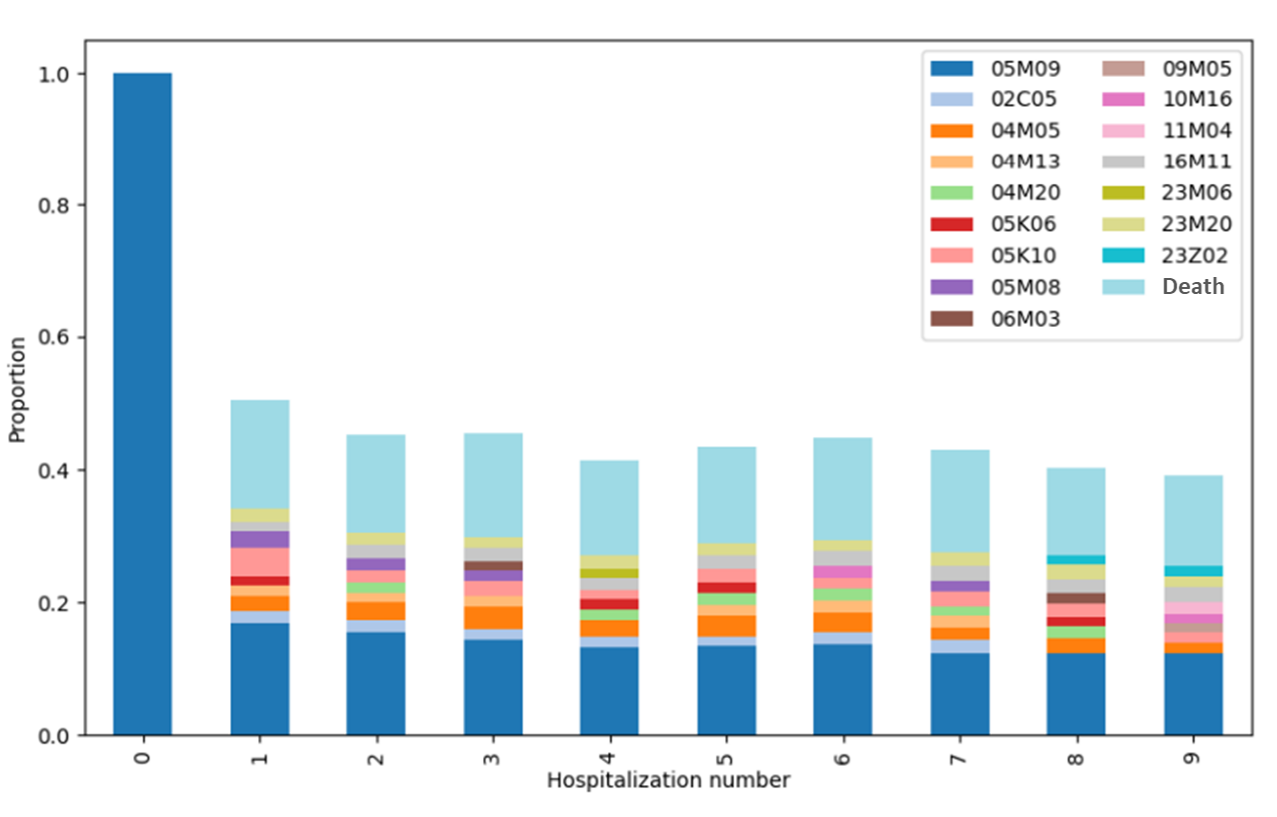}  
    \caption{Frequent ICD-10 codes}
\end{subfigure}
\begin{subfigure}{.45\textwidth}
    \centering
    \includegraphics[width=\linewidth]{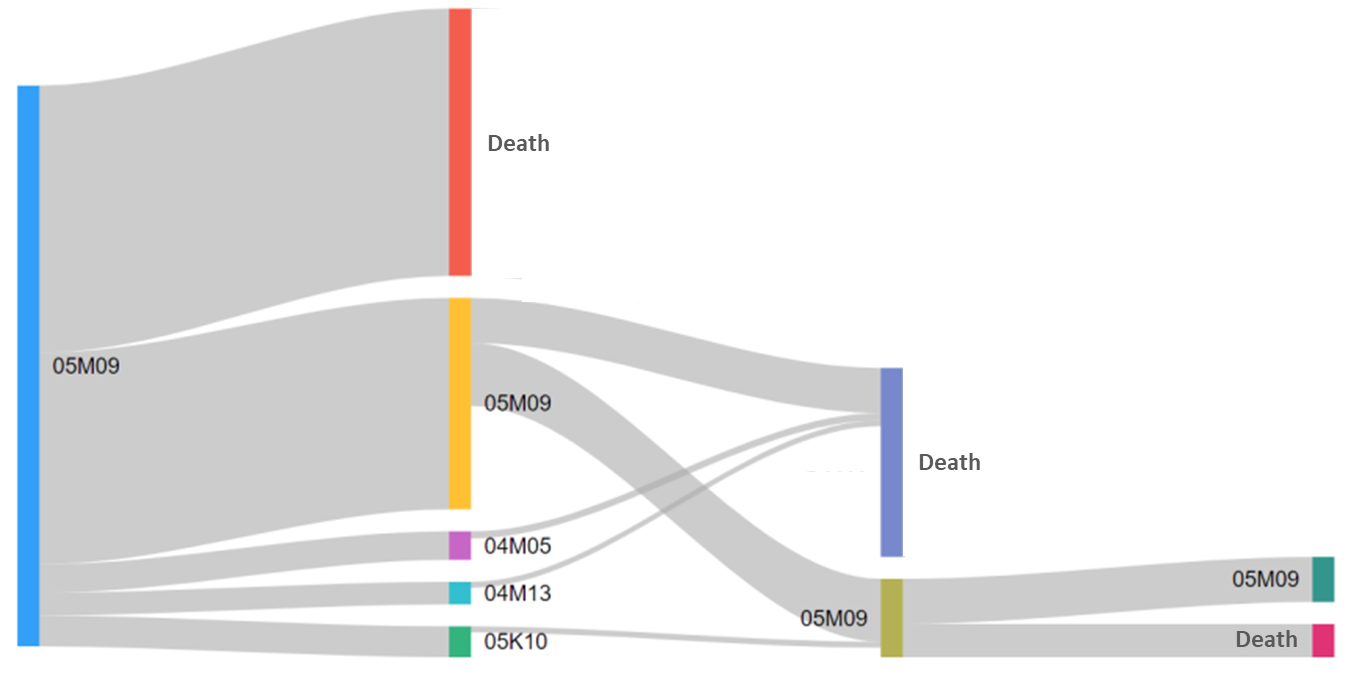}  
    \caption{Sankey diagram}
\end{subfigure}
\caption{Key figures for Cluster 1}
    \label{fig:clust1}
\end{figure}

When we examined the ten most frequent ICD-10 codes for the $n^{th}$ hospitalization after HF, we observed that these ICD-10 codes collectively accounted for approximately 50\% from the first to the tenth occurrence. This indicates significant similarity in the care sequences of these patients (Figure \ref{fig:clust1}).  Furthermore, by visualizing the frequent trajectories for the entire HF patients population, we can identify several common sequences leading to death (e.g., '05M09' for HF hospitalization, '04M05' for pleurisy, or '04M13' for pulmonary edema and respiratory distress).

We also observed a high proportion of deaths during the initial hospitalizations in clusters 2 and 4, consistent with the older age and shorter hospitalization sequences of these patients (see Appendix). Additionally, HF patients in cluster 2 who experienced mortality had no more than six hospitalizations following their first HF episode. 

\begin{figure}[h]
    \centering
    \includegraphics[width=0.55\linewidth]{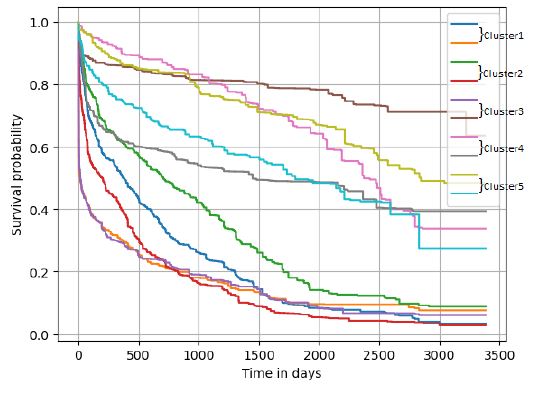}
    \caption{Survival predictions in clusters with best and worst scenarios}
    \label{fig:surv-clust}
\end{figure}

Figure \ref{fig:surv-clust} displays the survival trajectories for the most and least optimistic scenarios within each cluster. Since clusters have varying numbers of individuals, prediction accuracy varies considerably across models. Specifically, cluster 3 yields more uncertain predictions compared to cluster 5. Aging consistently emerged as a significant factor contributing to mortality across all clusters, as well as being male. Prolonged hospital stays are also associated with a more pessimistic trajectory. We obtained similar results using RSF and survival gradient boosting, yielding a mean C-index of 0.68 (details in Appendix).

\section{Discussion}
By systematically identifying and analyzing frequent hospitalization patterns, we gained valuable insights into the care sequences of heart failure patients. This comprehensive analysis enabled to not only discern the most common reasons for hospitalization but also to trace the trajectories that often culminate in patient mortality. 

Our study benefits from a large dataset, which provides a solid foundation for our analyses. A significant highlight of our methodology is the use of unsupervised clustering, not least with the introduction of a novel distance metric. More recently, methodological research has been conducted to create similarity measures for clustering purposes, either based on prevalence \citep{mannino2017development}, or Bayesian framework \citep{wu2020maintenance}. However, our approach enables to keep into account the numbered order of hospitalizations, and to assess the impact on death using the survival analysis \citep{pinaire2017patient}. Our methodology also allows us to avoid making any \textit{a priori} assumptions about the patient population, thereby reducing potential biases associated with specific patient characteristics (clinical discussion is available in Appendix \ref{app:cd}).

Health trajectories is a burning topic in clinical research. Within our methodological framework, we introduce an innovative approach to comprehending hospitalization sequences and their implications for survival outcomes. While our focus has been on the HF patients, our approach is adaptable and can be extended to address more intricate populations, and this way to meet a variety of clinical challenges.

\bibliography{iclr2024_conference}

\begin{thebibliography}{34}
\providecommand{\natexlab}[1]{#1}
\providecommand{\url}[1]{\texttt{#1}}
\expandafter\ifx\csname urlstyle\endcsname\relax
  \providecommand{\doi}[1]{doi: #1}\else
  \providecommand{\doi}{doi: \begingroup \urlstyle{rm}\Url}\fi

\bibitem[Akiba et~al.(2019)Akiba, Sano, Yanase, Ohta, and Koyama]{akiba2019optuna}
Takuya Akiba, Shotaro Sano, Toshihiko Yanase, Takeru Ohta, and Masanori Koyama.
\newblock Optuna: A next-generation hyperparameter optimization framework.
\newblock In \emph{Proceedings of the 25th ACM SIGKDD international conference on knowledge discovery \& data mining}, pp.\  2623--2631, 2019.

\bibitem[Al-Maolegi \& Arkok(2014)Al-Maolegi and Arkok]{al-maolegi_improved_2014}
Mohammed Al-Maolegi and Bassam Arkok.
\newblock An {Improved} {Apriori} {Algorithm} {For} {Association} {Rules}.
\newblock \emph{International Journal on Natural Language Computing}, 3\penalty0 (1):\penalty0 21--29, February 2014.
\newblock ISSN 23194111, 22781307.
\newblock \doi{10.5121/ijnlc.2014.3103}.
\newblock URL \url{http://www.airccse.org/journal/ijnlc/papers/3114ijnlc03.pdf}.

\bibitem[Choua{\"\i}d et~al.(2022)Choua{\"\i}d, Grumberg, Batisse, Corre, Giaj~Levra, Gaudin, Prodel, Lortet-Tieulent, Assi{\'e}, and Cott{\'e}]{chouaid2022machine}
Christos Choua{\"\i}d, Valentine Grumberg, Alexandre Batisse, Romain Corre, Matteo Giaj~Levra, Anne-Fran{\c{c}}oise Gaudin, Martin Prodel, Joannie Lortet-Tieulent, Jean-Baptiste Assi{\'e}, and Francois-Emery Cott{\'e}.
\newblock Machine learning--based analysis of treatment sequences typology in advanced non--small-cell lung cancer long-term survivors treated with nivolumab.
\newblock \emph{JCO Clinical Cancer Informatics}, 6:\penalty0 e2100108, 2022.

\bibitem[Constantinou et~al.(2021)Constantinou, Pelletier-Fleury, Oli{\'e}, Gastaldi-M{\'e}nager, Juill{\`E}re, and Tuppin]{constantinou2021patient}
Panayotis Constantinou, Nathalie Pelletier-Fleury, Val{\'e}rie Oli{\'e}, Christelle Gastaldi-M{\'e}nager, Yves Juill{\`E}re, and Philippe Tuppin.
\newblock Patient stratification for risk of readmission due to heart failure by using nationwide administrative data.
\newblock \emph{Journal of Cardiac Failure}, 27\penalty0 (3):\penalty0 266--276, 2021.

\bibitem[Cox(1972)]{cox1972regression}
David~R Cox.
\newblock Regression models and life-tables.
\newblock \emph{Journal of the Royal Statistical Society: Series B (Methodological)}, 34\penalty0 (2):\penalty0 187--202, 1972.

\bibitem[De~Roquefeuil et~al.(2009)De~Roquefeuil, Studer, Neumann, and Merlière]{de_roquefeuil_lechantillon_2009}
Laurence De~Roquefeuil, A~Studer, A.~Neumann, and Y.~Merlière.
\newblock L'échantillon généraliste de bénéficiaires : représentativité, portée et limites.
\newblock \emph{Pratiques et Organisation des Soins}, 40\penalty0 (3):\penalty0 213--223, 2009.
\newblock ISSN 1952-9201.
\newblock \doi{10.3917/pos.403.0213}.
\newblock URL \url{https://www.cairn.info/revue-pratiques-et-organisation-des-soins-2009-3-page-213.htm}.

\bibitem[Farré et~al.(2017)Farré, Vela, Clèries, Bustins, Cainzos-Achirica, Enjuanes, Moliner, Ruiz, Verdú-Rotellar, and Comín-Colet]{farre_real_2017}
Núria Farré, Emili Vela, Montse Clèries, Montse Bustins, Miguel Cainzos-Achirica, Cristina Enjuanes, Pedro Moliner, Sonia Ruiz, José~María Verdú-Rotellar, and Josep Comín-Colet.
\newblock Real world heart failure epidemiology and outcome: {A} population-based analysis of 88,195 patients.
\newblock \emph{PLOS ONE}, 12\penalty0 (2):\penalty0 e0172745, February 2017.
\newblock ISSN 1932-6203.
\newblock \doi{10.1371/journal.pone.0172745}.
\newblock URL \url{https://dx.plos.org/10.1371/journal.pone.0172745}.

\bibitem[Harrell et~al.(1982)Harrell, Califf, Pryor, Lee, and Rosati]{harrell1982evaluating}
Frank~E Harrell, Robert~M Califf, David~B Pryor, Kerry~L Lee, and Robert~A Rosati.
\newblock Evaluating the yield of medical tests.
\newblock \emph{Jama}, 247\penalty0 (18):\penalty0 2543--2546, 1982.

\bibitem[Harrell et~al.(1984)Harrell, Lee, Califf, Pryor, and Rosati]{harrell1984regression}
Frank~E Harrell, Kerry~L Lee, Robert~M Califf, David~B Pryor, and Robert~A Rosati.
\newblock Regression modelling strategies for improved prognostic prediction.
\newblock \emph{Statistics in medicine}, 3\penalty0 (2):\penalty0 143--152, 1984.

\bibitem[Hewamalage et~al.(2021)Hewamalage, Bergmeir, and Bandara]{hewamalage2021recurrent}
Hansika Hewamalage, Christoph Bergmeir, and Kasun Bandara.
\newblock Recurrent neural networks for time series forecasting: Current status and future directions.
\newblock \emph{International Journal of Forecasting}, 37\penalty0 (1):\penalty0 388--427, 2021.

\bibitem[Hothorn et~al.(2006)Hothorn, B{\"u}hlmann, Dudoit, Molinaro, and Van Der~Laan]{hothorn2006survival}
Torsten Hothorn, Peter B{\"u}hlmann, Sandrine Dudoit, Annette Molinaro, and Mark~J Van Der~Laan.
\newblock Survival ensembles.
\newblock \emph{Biostatistics}, 7\penalty0 (3):\penalty0 355--373, 2006.

\bibitem[Hu(2007)]{hu2007akaike}
Shuhua Hu.
\newblock Akaike information criterion.
\newblock \emph{Center for Research in Scientific Computation}, 93:\penalty0 42, 2007.

\bibitem[Ishwaran et~al.(2008)Ishwaran, Kogalur, Blackstone, and Lauer]{ishwaran_random_2008}
Hemant Ishwaran, Udaya~B. Kogalur, Eugene~H. Blackstone, and Michael~S. Lauer.
\newblock Random survival forests.
\newblock \emph{The Annals of Applied Statistics}, 2\penalty0 (3):\penalty0 841--860, September 2008.
\newblock ISSN 1932-6157, 1941-7330.
\newblock \doi{10.1214/08-AOAS169}.
\newblock URL \url{https://projecteuclid.org/journals/annals-of-applied-statistics/volume-2/issue-3/Random-survival-forests/10.1214/08-AOAS169.full}.

\bibitem[{Jian Pei} et~al.(2001){Jian Pei}, {Jiawei Han}, Mortazavi-Asl, Pinto, {Qiming Chen}, Dayal, and {Mei-Chun Hsu}]{jian_pei_prefixspan_2001}
{Jian Pei}, {Jiawei Han}, B.~Mortazavi-Asl, H.~Pinto, {Qiming Chen}, U.~Dayal, and {Mei-Chun Hsu}.
\newblock {PrefixSpan},: mining sequential patterns efficiently by prefix-projected pattern growth.
\newblock In \emph{Proceedings 17th {International} {Conference} on {Data} {Engineering}}, pp.\  215--224, Heidelberg, Germany, 2001. IEEE Comput. Soc.
\newblock ISBN 978-0-7695-1001-9.
\newblock \doi{10.1109/ICDE.2001.914830}.
\newblock URL \url{http://ieeexplore.ieee.org/document/914830/}.

\bibitem[Kang et~al.(2020)Kang, Baek, and Chung]{kang2020prefixspan}
Ji-Soo Kang, Ji-Won Baek, and Kyungyong Chung.
\newblock Prefixspan based pattern mining using time sliding weight from streaming data.
\newblock \emph{IEEE Access}, 8:\penalty0 124833--124844, 2020.

\bibitem[Kaur et~al.(2014)Kaur, Kaur, and Singh]{kaur2014k}
Noor~Kamal Kaur, Usvir Kaur, and Dheerendra Singh.
\newblock K-medoid clustering algorithm-a review.
\newblock \emph{Int. J. Comput. Appl. Technol}, 1\penalty0 (1):\penalty0 42--45, 2014.

\bibitem[Lambert et~al.(2023)Lambert, Leutenegger, Jannot, and Baudot]{lambert2023tracking}
Judith Lambert, Anne-Louise Leutenegger, Anne-Sophie Jannot, and Ana{\"\i}s Baudot.
\newblock Tracking clusters of patients over time enables extracting information from medico-administrative databases.
\newblock \emph{Journal of Biomedical Informatics}, 139:\penalty0 104309, 2023.

\bibitem[Leis et~al.(2023)Leis, McSpadden, Segaloff, Lauring, Cheng, Petrie, Lamerato, Patel, Flannery, Ferdinands, et~al.]{leis2023k}
Aleda~M Leis, Erin McSpadden, Hannah~E Segaloff, Adam~S Lauring, Caroline Cheng, Joshua~G Petrie, Lois~E Lamerato, Manish Patel, Brendan Flannery, Jill Ferdinands, et~al.
\newblock K-medoids clustering of hospital admission characteristics to classify severity of influenza virus infection.
\newblock \emph{Influenza and Other Respiratory Viruses}, 17\penalty0 (3):\penalty0 e13120, 2023.

\bibitem[Mabroukeh \& Ezeife(2010)Mabroukeh and Ezeife]{mabroukeh2010taxonomy}
Nizar~R Mabroukeh and Christie~I Ezeife.
\newblock A taxonomy of sequential pattern mining algorithms.
\newblock \emph{ACM Computing Surveys (CSUR)}, 43\penalty0 (1):\penalty0 1--41, 2010.

\bibitem[MacDonald \& Zucchini(1997)MacDonald and Zucchini]{macdonald1997hidden}
Iain~L MacDonald and Walter Zucchini.
\newblock \emph{Hidden Markov and other models for discrete-valued time series}, volume 110.
\newblock CRC Press, 1997.

\bibitem[Mannino et~al.(2017)Mannino, Fredrickson, Banaei-Kashani, Linck, and Raghda]{mannino2017development}
Michael Mannino, Joel Fredrickson, Farnoush Banaei-Kashani, Iris Linck, and Raghda~Alqurashi Raghda.
\newblock Development and evaluation of a similarity measure for medical event sequences.
\newblock \emph{ACM Transactions on Management Information Systems (TMIS)}, 8\penalty0 (2-3):\penalty0 1--26, 2017.

\bibitem[Masseglia et~al.(2004)Masseglia, Teisseire, and Poncelet]{masseglia_extraction_2004}
Florent Masseglia, Maguelonne Teisseire, and Pascal Poncelet.
\newblock Extraction de motifs séquentiels. {Problèmes} et méthodes.
\newblock \emph{Revue des Sciences et Technologies de l'Information - Série ISI : Ingénierie des Systèmes d'Information}, 9\penalty0 (3/4):\penalty0 183, 2004.
\newblock \doi{10.3166/isi.9.3-4.183-210}.
\newblock URL \url{https://hal-lirmm.ccsd.cnrs.fr/lirmm-00108563}.

\bibitem[Murris et~al.(2023)Murris, Charles-Nelson, Tadmouri~Sellier, Lavenu, and Katsahian]{murris2023towards}
Juliette Murris, Anais Charles-Nelson, Abir Tadmouri~Sellier, Audrey Lavenu, and Sandrine Katsahian.
\newblock Towards filling the gaps around recurrent events in high dimensional framework: a systematic literature review and application.
\newblock \emph{Biostatistics \& Epidemiology}, 7\penalty0 (1):\penalty0 e2283650, 2023.

\bibitem[Nguyen et~al.(2018)Nguyen, Luo, Venkatesh, and Phung]{nguyen2018effective}
Dang Nguyen, Wei Luo, Svetha Venkatesh, and Dinh Phung.
\newblock Effective identification of similar patients through sequential matching over icd code embedding.
\newblock \emph{Journal of medical systems}, 42:\penalty0 1--13, 2018.

\bibitem[Pei et~al.(2001)Pei, Han, Mortazavi-Asl, Pinto, Chen, Dayal, and Hsu]{pei_prefixspan_2001}
Jian Pei, Jiawei Han, B.~Mortazavi-Asl, H.~Pinto, Qiming Chen, U.~Dayal, and Mei-Chun Hsu.
\newblock {PrefixSpan},: mining sequential patterns efficiently by prefix-projected pattern growth.
\newblock In \emph{Proceedings 17th {International} {Conference} on {Data} {Engineering}}, pp.\  215--224, April 2001.
\newblock \doi{10.1109/ICDE.2001.914830}.
\newblock ISSN: 1063-6382.

\bibitem[Pinaire(2017)]{pinaire_explorer_2017}
Jessica Pinaire.
\newblock \emph{Explorer les trajectoires de patients via les bases médico-économiques : application à l'infarctus du myocarde}.
\newblock phdthesis, Université Montpellier, October 2017.
\newblock URL \url{https://theses.hal.science/tel-01903477}.

\bibitem[Pinaire et~al.(2017)Pinaire, Az{\'e}, Bringay, and Landais]{pinaire2017patient}
Jessica Pinaire, J{\'e}r{\^o}me Az{\'e}, Sandra Bringay, and Paul Landais.
\newblock Patient healthcare trajectory. an essential monitoring tool: a systematic review.
\newblock \emph{Health information science and systems}, 5:\penalty0 1--18, 2017.

\bibitem[Savarese et~al.(2022)Savarese, Becher, Lund, Seferovic, Rosano, and Coats]{savarese2022global}
Gianluigi Savarese, Peter~Moritz Becher, Lars~H Lund, Petar Seferovic, Giuseppe~MC Rosano, and Andrew~JS Coats.
\newblock Global burden of heart failure: a comprehensive and updated review of epidemiology.
\newblock \emph{Cardiovascular research}, 118\penalty0 (17):\penalty0 3272--3287, 2022.

\bibitem[Simpson et~al.(2020)Simpson, Jhund, Lund, Padmanabhan, Claggett, Shen, Petrie, Abraham, Desai, Dickstein, et~al.]{simpson2020prognostic}
Joanne Simpson, Pardeep~S Jhund, Lars~H Lund, Sandosh Padmanabhan, Brian~L Claggett, LI~Shen, Mark~C Petrie, William~T Abraham, Akshay~S Desai, Kenneth Dickstein, et~al.
\newblock Prognostic models derived in paradigm-hf and validated in atmosphere and the swedish heart failure registry to predict mortality and morbidity in chronic heart failure.
\newblock \emph{JAMA cardiology}, 5\penalty0 (4):\penalty0 432--441, 2020.

\bibitem[Tibshirani(1996)]{tibshirani_regression_1996}
Robert Tibshirani.
\newblock Regression {Shrinkage} and {Selection} {Via} the {Lasso}.
\newblock \emph{Journal of the Royal Statistical Society: Series B (Methodological)}, 58\penalty0 (1):\penalty0 267--288, January 1996.
\newblock ISSN 00359246.
\newblock \doi{10.1111/j.2517-6161.1996.tb02080.x}.
\newblock URL \url{https://onlinelibrary.wiley.com/doi/10.1111/j.2517-6161.1996.tb02080.x}.

\bibitem[Wu \& Hao(2020)Wu and Hao]{wu2020maintenance}
Zhenya Wu and Jianping Hao.
\newblock A maintenance task similarity-based prior elicitation method for bayesian maintainability demonstration.
\newblock \emph{Mathematical Problems in Engineering}, 2020:\penalty0 1--19, 2020.

\bibitem[Yujian \& Bo(2007)Yujian and Bo]{yujian2007normalized}
Li~Yujian and Liu Bo.
\newblock A normalized levenshtein distance metric.
\newblock \emph{IEEE transactions on pattern analysis and machine intelligence}, 29\penalty0 (6):\penalty0 1091--1095, 2007.

\bibitem[Zhong et~al.(2021)Zhong, Loukides, and Pissis]{zhong2021clustering}
Haodi Zhong, Grigorios Loukides, and Solon~P Pissis.
\newblock Clustering demographics and sequences of diagnosis codes.
\newblock \emph{IEEE Journal of Biomedical and Health Informatics}, 26\penalty0 (5):\penalty0 2351--2359, 2021.

\bibitem[Zhu et~al.(2021)Zhu, Plasek, Tang, Al-Assad, Zhang, Xiong, Wang, Yerneni, Ortega, Kang, et~al.]{zhu2021embedding}
Xudong Zhu, Joseph~M Plasek, Chunlei Tang, Wasim Al-Assad, Zhikun Zhang, Yun Xiong, Liqin Wang, Sharmitha Yerneni, Carlos Ortega, Min-Jeoung Kang, et~al.
\newblock Embedding, aligning and reconstructing clinical notes to explore sepsis.
\newblock \emph{BMC Research Notes}, 14:\penalty0 1--6, 2021.

\end{thebibliography}
\bibliographystyle{iclr2024_conference}

\newpage
\appendix
\section{Appendix}
\subsection{Sequential pattern mining}
\subsubsection{Definitions}
\textbf{Itemset.} Let $I = i_1,...,i_{N_p}$ be the set of $N_p$ items. A subset of $I$ is called an itemset. In this study, a pattern or itemset consists in an ICD-10 code.

\textbf{Event sequence.} An event sequence $seq = \{e_1,...,e_m\}  , e_i \subseteq I$ for $1 \leq i \leq m$ is an ordered list of itemsets. The event sequence database is the starting point for sequential pattern mining. 

\textbf{Subsequence.} An event sequence $seq_{sub} = \{r_1,...,r_q\}$ is a subsequence of $seq$ if there exist integers $1 \leq i_1 \leq... \leq i_q \leq m$, s. t. $r_1 \subseteq e_{i_1}, ..., r_q \subseteq e_{i_q}$.

\textbf{Support.} Let $B = \{seq_1, ..., seq_{N_p}\}$ a set of sequences. The support $Freq_B(seq)$ of a sequence $seq$ is the number of sequences in $B$ that have $seq$ as a subsequence. The higher the support, the more frequently the pattern occurs in the database.

\textbf{Frequent sequential pattern.} An event sequence $seq$ is frequent and called a frequent sequential pattern if its support is greater than or equal to a minimum threshold $\sigma: Freq_B(seq) \geq \sigma$.

\subsubsection{Algorithm}
The \textit{PrefixSpan} algorithm was retained and uses the concept of "prefixes" to efficiently search for frequent patterns in a sequence database . The algorithm works by first identifying all frequent single-item sequences, and then iteratively extending these prefixes to form longer sequential patterns. 

The algorithm starts with an empty prefix and the entire dataset as the initial projected database. It then recursively explores and extends the prefixes while checking the support of the generated sequences. Frequent sequences above the minimum support threshold are output, and the process continues until no more frequent sequences can be found. 
\begin{algorithm}
    \caption{The \textit{PrefixSpan} algorithm}\label{alg:cap}
    \begin{algorithmic}
        \item 
            \begin{enumerate}
                \item \textbf{Initialize}: Start with an empty set of frequent patterns $I_0 = \emptyset$
                \item \textbf{Frequent Items}: For each item $i$ in the first sequence $seq_1$ of the input data, create a singleton pattern
                \item \textbf{Generate Sequences}: For each frequent item $i$ found in step 2, extend the current prefix sequence by adding that item to the set of frequent patterns
                \item \textbf{Recursive Search}: For each new sequence created in step 3, repeat steps 2 and 3 recursively. For each pattern:
                \begin{enumerate}
                    \item Construct a database of all sequences that contain the pattern as a subsequence
                    \item For each item that appears after the last item of the pattern in the input data, create a new pattern by extending the pattern with the item
                    \item Compute the support of the new pattern by concatenating the support of the item with the support of the database
                    \item If the new pattern is frequent in the database, add it to the set of frequent patterns and continue the recursive search
                \end{enumerate}
            \end{enumerate} 
    \end{algorithmic}
\end{algorithm}

\textit{PrefixSpan} algorithm is known for its efficiency and scalability, particularly for mining long sequential patterns \citep{mabroukeh2010taxonomy}. Of note, several sequential pattern mining algorithms were experimented, like \textit{APriori} \citep{al-maolegi_improved_2014}. Similar support for patterns were found but computing time was much higher. This is in line with existing literature in terms of run-time and memory usage \citep{jian_pei_prefixspan_2001}.

\begin{table}[H]
\caption{Most occurring patterns in health care pathway for each patient clusters}
\label{tab:top1_cluster}
\resizebox{\textwidth}{!}{%
\renewcommand{\arraystretch}{2}
\begin{tabular}{lrrlrrlrrl}
\toprule
{\textbf{Cluster}} &  \textbf{Count} &  \textbf{Freq.}& \textbf{Top1 pattern (len1)} &  \textbf{Count} &  \textbf{Freq.} & \textbf{Top1 pattern (len2)} &   \textbf{Count}&  \textbf{Freq.} & \textbf{Top1 pattern (len3)} \\
\midrule
1 & 833 & 0.610 & ['Death'] & 507 & 0.371 & ['05M09', 'Death'] & 301 & 0.220 & ['05M09', '05M09', 'Death'] \\
2 & 3467 & 0.685 & ['Death'] & 1542 & 0.305 & ['05M09', 'Death'] & 336 & 0.066 & ['05M09', '05M09', 'Death'] \\
3 & 22 & 0.629 & ['23M20'] & 15 & 0.429 & ['23M20', '23M20'] & 8 & 0.229 & ['23M20', '23M20', '23M20'] \\
4 & 2082 & 0.651 & ['Death'] & 1111 & 0.347 & ['05M09', 'Death'] & 487 & 0.152 & ['05M09', '05M09', 'Death'] \\
5 & 224 & 0.579 & ['05M09'] & 132 & 0.341 & ['05M09', '05M09'] & 75 & 0.194 & ['05M09', '05M09', '05M09'] \\
\bottomrule
\end{tabular}}
\end{table}

\begin{table}[H]
\caption{Second most occurring patterns in health care pathway for each patient clusters}
\label{tab:top2_cluster}
\resizebox{\textwidth}{!}{%
\renewcommand{\arraystretch}{2}
\begin{tabular}{lrrlrrlrrl}
\toprule
{\textbf{Cluster}} &  \textbf{Count} &  \textbf{Freq.}& \textbf{Top2 pattern (len1)} &  \textbf{Count} &  \textbf{Freq.} & \textbf{Top2 pattern (len2)} &   \textbf{Count}&  \textbf{Freq.} & \textbf{Top2 pattern (len3)} \\
\midrule
1 & 777 & 0.569 & ['05M09'] & 431 & 0.316 & ['05M09', '05M09'] & 256 & 0.187 & ['05M09', '05M09', '05M09'] \\
2 & 2032 & 0.401 & ['05M09'] & 489 & 0.097 & ['\textbf{04M05}', 'Death'] & 104 & 0.021 & ['05M09', '05M09', '05M09'] \\
3 & 18 & 0.514 & ['05M09'] & 10 & 0.286 & ['23M20', '16M11'] & 8 & 0.229 & ['23M20', '23M20', '23M20'] \\
4 & 1586 & 0.496 & ['05M09'] & 642 & 0.201 & ['05M09', '05M09'] & 285 & 0.089 & ['05M09', '05M09', '05M09'] \\
5 & 223 & 0.576 & ['Death'] & 129 & 0.333 & ['05M09', 'Death'] & 75 & 0.194 & ['05M09', '05M09', 'Death'] \\
\bottomrule
\end{tabular}}
\end{table}

\begin{table}[H]
\caption{Third most occurring patterns in health care pathway for each patient clusters}
\label{tab:top3_cluster}
\resizebox{\textwidth}{!}{%
\renewcommand{\arraystretch}{2}
\begin{tabular}{lrrlrrlrrl}
\toprule
{\textbf{Cluster}} &  \textbf{Count} &  \textbf{Freq.}& \textbf{Top3 pattern (len1)} &  \textbf{Count} &  \textbf{Freq.} & \textbf{Top3 pattern (len2)} &   \textbf{Count}&  \textbf{Freq.} & \textbf{Top3 pattern (len3)} \\
\midrule
1 & 456 & 0.334 & ['\textbf{05K10}'] & 255 & 0.187 & ['\textbf{04M05}', 'Death'] & 124 & 0.091 & ['\textbf{05K10}', '05M09', 'Death'] \\
2 & 615 & 0.121 & ['\textbf{04M05}'] & 418 & 0.083 & ['05M09', '05M09'] & 102 & 0.020 & ['02C05', '05M09', 'Death'] \\
3 & 13 & 0.371 & ['06M03'] & 9 & 0.257 & ['23M20', '05M09'] & 8 & 0.229 & ['23M20', '23M20', '23M20'] \\
4 & 823 & 0.257 & ['02C05'] & 505 & 0.158 & ['\textbf{04M05}', 'Death'] & 208 & 0.065 & ['02C05', '05M09', 'Death'] \\
5 & 167 & 0.432 & ['\textbf{05K10}'] & 88 & 0.227 & ['05K10', '05M09'] & 56 & 0.145 & ['05K10', '05M09', '05M09'] \\
\bottomrule
\end{tabular}}
\end{table}

\newpage
\begin{table}[H]
\caption{Proportions of top 10 ICD-10 codes in nth hospitalization after first hospitalization for heart failure - Deceased
patients}
\label{tab:freq_GHMs}
\resizebox{\textwidth}{!}{%
\renewcommand{\arraystretch}{1}
\begin{tabular}{lllllllllll}
\toprule
& 0 & 1 & 2 & 3 & 4 & 5 & 6 & 7 & 8 & 9 \\
\midrule
\textbf{05M09} & 1.000 & 0.169 & 0.155 & 0.143 & 0.131 & 0.134 & 0.137 & 0.124 & 0.124 & 0.123 \\
\textbf{Death} &  & 0.165 & 0.148 & 0.156 & 0.143 & 0.144 & 0.154 & 0.156 & 0.132 & 0.137 \\
\textbf{05K10} &  & \textcolor{red}{0.043} & 0.020 & \textcolor{brown}{0.021} & 0.014 & 0.019 & 0.017 & \textcolor{red}{0.023} & \textcolor{brown}{0.022} & 0.015 \\
\textbf{05M08} &  & \textcolor{brown}{0.024} & 0.017 & 0.016 &  &  &  & 0.016 &  &  \\
\textbf{04M05} &  & 0.023 & \textcolor{red}{0.028} & \textcolor{red}{0.033} & \textcolor{red}{0.026} & \textcolor{red}{0.030} & \textcolor{red}{0.029} & 0.018 & \textcolor{red}{0.023} & 0.015 \\
\textbf{23M20} &  & 0.019 & 0.019 & 0.016 & \textcolor{brown}{0.022} & 0.018 & 0.017 & \textcolor{brown}{0.020} & \textcolor{brown}{0.022} & 0.015 \\
\textbf{04M13} &  & 0.017 & 0.014 & 0.016 &  & 0.016 & 0.019 & 0.018 &  &  \\
\textbf{02C05} &  & 0.016 & 0.017 & 0.017 & 0.016 & 0.014 & 0.019 & \textcolor{brown}{0.020} &  &  \\
\textbf{16M11} &  & 0.015 & \textcolor{brown}{0.020} & 0.020 & 0.019 & \textcolor{brown}{0.023} & \textcolor{brown}{0.021} & \textcolor{red}{0.023} & \textcolor{brown}{0.022} & \textcolor{red}{0.022} \\
\textbf{05K06} &  & 0.013 &  &  & 0.015 & 0.016 &  &  & 0.014 &  \\
\textbf{04M20} &  &  & 0.014 &  & 0.016 & 0.019 & 0.016 & 0.014 & 0.017 &  \\
\textbf{06M03} &  &  &  & 0.015 &  &  &  &  & 0.015 &  \\
\textbf{09M05} &  &  &  &  &  &  &  &  &  & 0.014 \\
\textbf{10M16} &  &  &  &  &  &  & 0.019 &  &  & 0.014 \\
\textbf{11M04} &  &  &  &  &  &  &  &  &  & \textcolor{brown}{0.018} \\
\textbf{23M06} &  &  &  &  & 0.012 &  &  &  &  &  \\
\textbf{23Z02} &  &  &  &  &  &  &  &  & 0.014 & 0.017\\
\bottomrule
\end{tabular}}
\end{table}

\newpage
\subsection{Clustering}
\subsubsection{Distance matrix}
Because we are working with strings of characters instead of numerical values - and we must keep those strings, we cannot encode them into numerical values - we cannot directly apply clustering algorithms on our data. With $N$ the total number of patients, the distance matrix $A$ writes
\begin{equation}
    \forall (i,j) \in [1,N], A_{i,j} = \mathcal{D}_P(patient_i,patient_j)
\end{equation}

Two properties are noted:
\begin{itemize}
    \item $\forall i \in [1,N], A_{i,i} = 0$,
    \item $A^T = A \Leftrightarrow \forall (i,j) \in [1,N], A_{i,j} = A_{j,i}$.
\end{itemize}

\subsubsection{K-medoids algorithm}
The algorithm proceeds iteratively:
\begin{enumerate}
    \item Select $k$ random medoids from the dataset;
    \item Each data point is assigned to its closest medoid and calculates the total distance between them;
    \item Improve the clustering by iteratively swapping one of the medoids with a non-medoid point and recompute the total distance;
    \item Whenever the total distance decreases, the swap is accepted and the new point becomes the medoid for the cluster.
\end{enumerate}
This process is repeated until no further improvement can be made. The pseudo-code is provided below.

\begin{algorithm}
\caption{K-Medoids Algorithm}\label{alg:kma}
\begin{algorithmic}[1]
\Require{$D$: dataset, $k$: number of clusters}
\Ensure{$C$: set of clusters, $M$: set of medoids}
\State Initialize $M$ with $k$ random data points from $D$
\State Assign each data point in $D$ to its closest medoid
\State Calculate the total distance $TD$ of all data points to their assigned medoids
\State $change \gets \textbf{true}$
\State $iter \gets 1$
\While{change}
\State $change \gets$ \textbf{false}
\ForAll{$m \in M$}
\ForAll{$p \in D \setminus M$}
\State Swap $m$ with $p$
\State Assign each data point in $D$ to its closest medoid
\State Calculate the total distance $TD'$ of all data points to their assigned medoids
\If{$TD' < TD$}
\State $M \gets$ updated set of medoids
\State $C \gets$ updated set of clusters
\State $TD \gets TD'$
\State $change \gets$ \textbf{true}
\Else
\State Swap $m$ with $p$ \Comment{Revert swap}
\EndIf
\EndFor
\EndFor
\EndWhile
\State \Return{$C$, $M$}
\end{algorithmic}
\end{algorithm}

\subsubsection{Clusters visualization}
Except for cluster 3, there were no significant differences amongst clusters. Cluster 3 with only 35 patients, stood out due to a considerably higher number and/or longer duration of hospitalizations compared to other patients. Other clusters included 1,366 patients (mean (sd) age = 78 (12.4) and 44.8\% were women), 5,063 patients (mean (sd) age = 83 (13.8) and 55.6\% were women), 3,200 patients (mean (sd) age = 81 (12.3) and 47.8\% were women), and 387 patients (mean (sd) age = 72 (13.8) and 44.2\% were women), respectively.

\begin{figure}[H]
\centering
\begin{subfigure}{.5\textwidth}
    \centering
    \includegraphics[width=\linewidth]{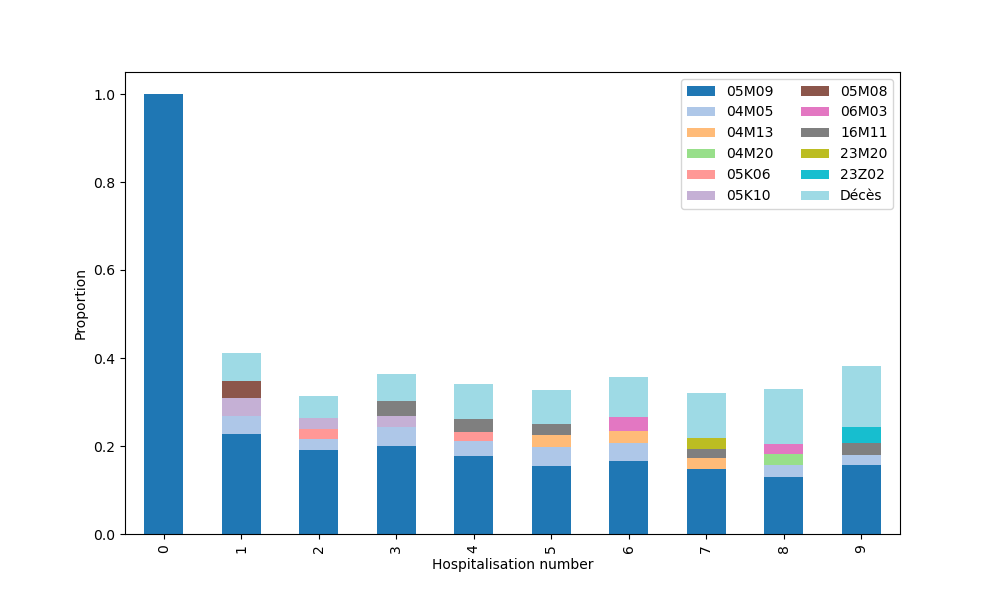}  
    \caption{Cluster 1}
\end{subfigure}
\begin{subfigure}{.5\textwidth}
    \centering
    \includegraphics[width=\linewidth]{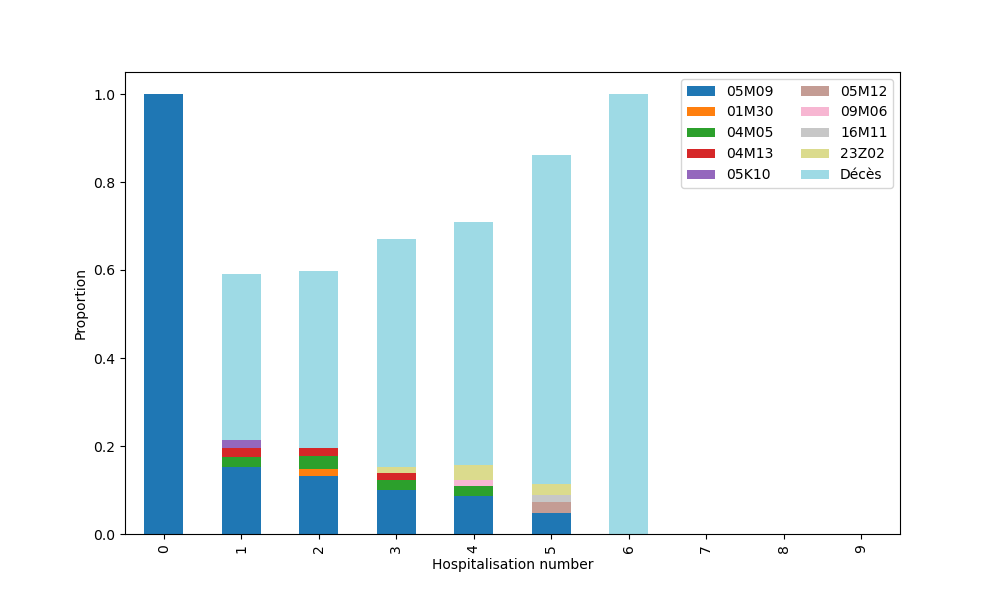}  
    \caption{Cluster 2}
\end{subfigure}
\begin{subfigure}{.49\textwidth}
    \centering
    \includegraphics[width=\linewidth]{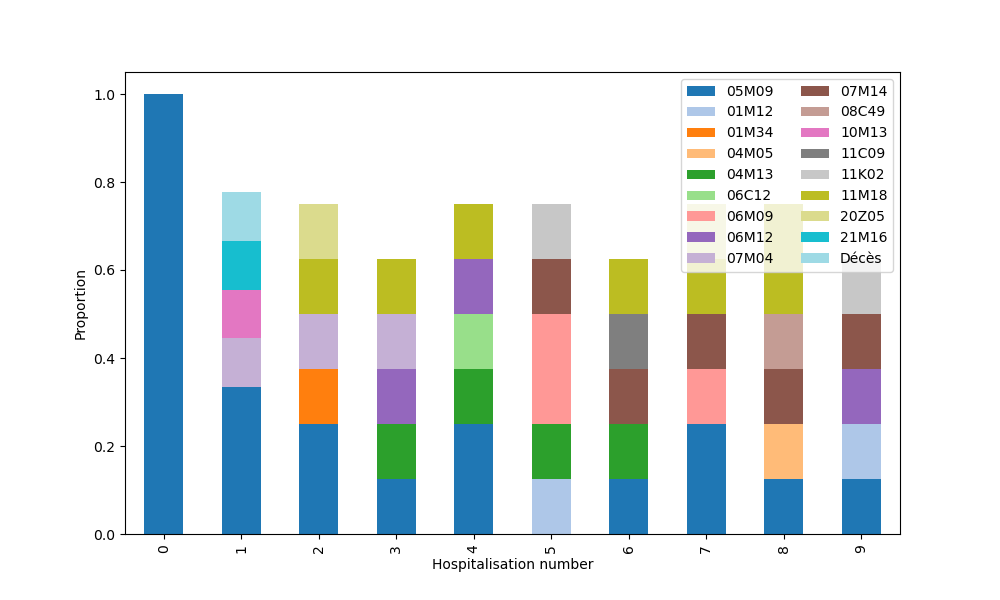}  
    \caption{Cluster 3}
\end{subfigure}
\begin{subfigure}{.49\textwidth}
    \centering
    \includegraphics[width=\linewidth]{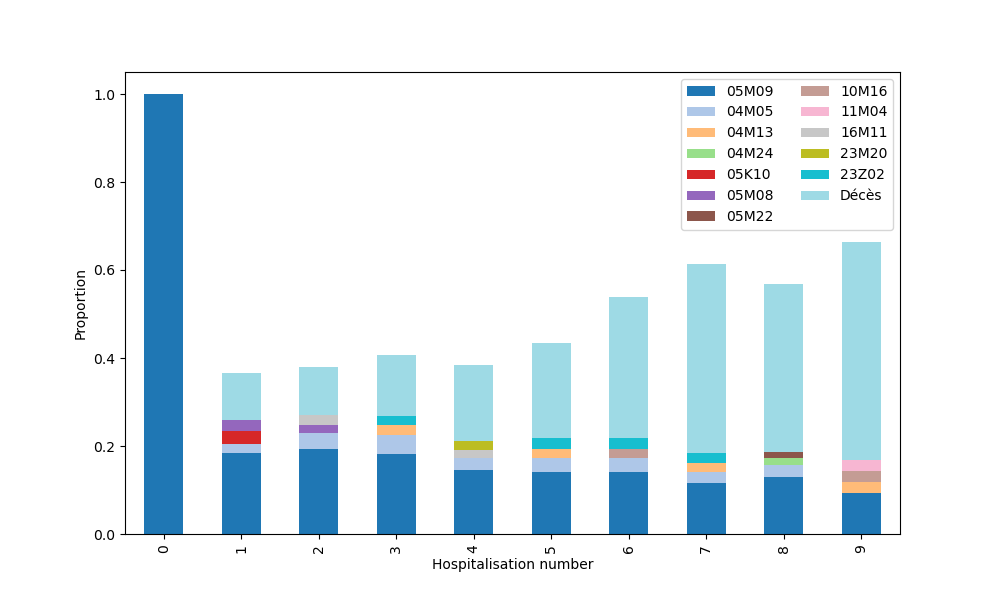}  
    \caption{Cluster 4}
\end{subfigure}
\begin{subfigure}{.49\textwidth}
    \centering
    \includegraphics[width=\linewidth]{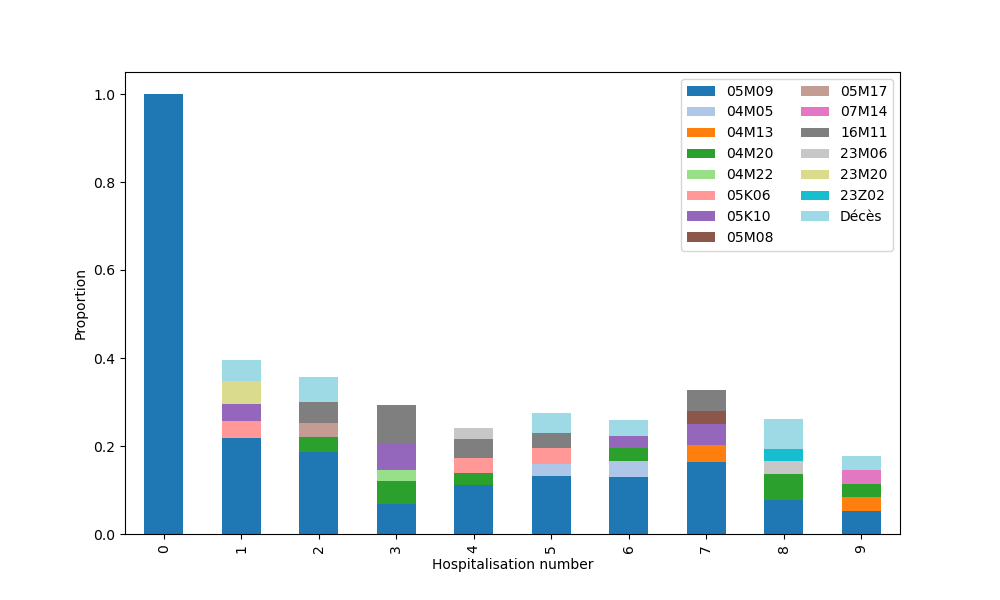}  
    \caption{Cluster 5}
\end{subfigure}
\caption{Most frequent ICD-10 codes after first hospitalization for heart failure - Deceased patients - Cluster 1 to 5}
\end{figure}

Here is the Python code to generate the figures above:
\begin{lstlisting}[language=Python, caption={Python code to compute most frequent ICD-10 codes after first hospitalization for heart failure }, label=python_code]
length_care_pathway = 10
top_k = 5 # we'll consider the top 5 most frequent ICD-10 codes 
          # for each step in the care pathways

database_top_k_frequent_codes = {}
for i in range(length_care_pathway):
    # for hospitalization i in all care pathways, 
    # consider the 5 most frequent ICD-10 codes found
    # the k most frequent codes per hospitalization step were found previoulsy 
    # through with text mining algorithms
    top_k_frequent_codes_i = df.iloc[i].sort_values(ascending=False)[:top_k]
    database_top_k_frequent_codes[f'{i}'] = top_k_frequent_codes_i

database_top_k_frequent_codes = pd.DataFrame(database_top_k_frequent_codes).transpose()
database_top_k_frequent_codes.plot(kind='bar', stacked=True));
\end{lstlisting}

\begin{table}[H] 
\caption{Samples of the database\_top\_k\_frequent\_codes dataframe}
\centering
\begin{adjustbox}{width=1.2\textwidth,center}
\begin{tabular}{lrrrrrrrrrrrrr}
\toprule
 & 04M05 & 04M13 & 04M24 & 05K10 & 05M08 & 05M09 & 05M22 & 10M16 & 11M04 & 16M11 & 23M20 & 23Z02 & Death \\
\midrule
0 & NaN & NaN & NaN & NaN & NaN & 1111.000000 & NaN & NaN & NaN & NaN & NaN & NaN \\
1 & 23.000000 & NaN & NaN & 34.000000 & 28.000000 & 204.000000 & NaN & NaN & NaN & NaN & NaN & 118.000000 \\
2 & 37.000000 & NaN & NaN & NaN & 18.000000 & 191.000000 & NaN & NaN & NaN & 23.000000 & NaN & NaN & 108.000000 \\
3 & 39.000000 & 20.000000 & NaN & NaN & NaN & 160.000000 & NaN & NaN & NaN & NaN & NaN & 18.000000 & 124.000000 \\
4 & 21.000000 & NaN & NaN & NaN & NaN & 110.000000 & NaN & NaN & NaN & 15.000000 & 14.000000 & NaN & 133.000000 \\
\bottomrule
\end{tabular}
\end{adjustbox}
\end{table}

\begin{figure}[H]
\centering
\begin{subfigure}{.5\textwidth}
    \centering
    \includegraphics[width=\linewidth]{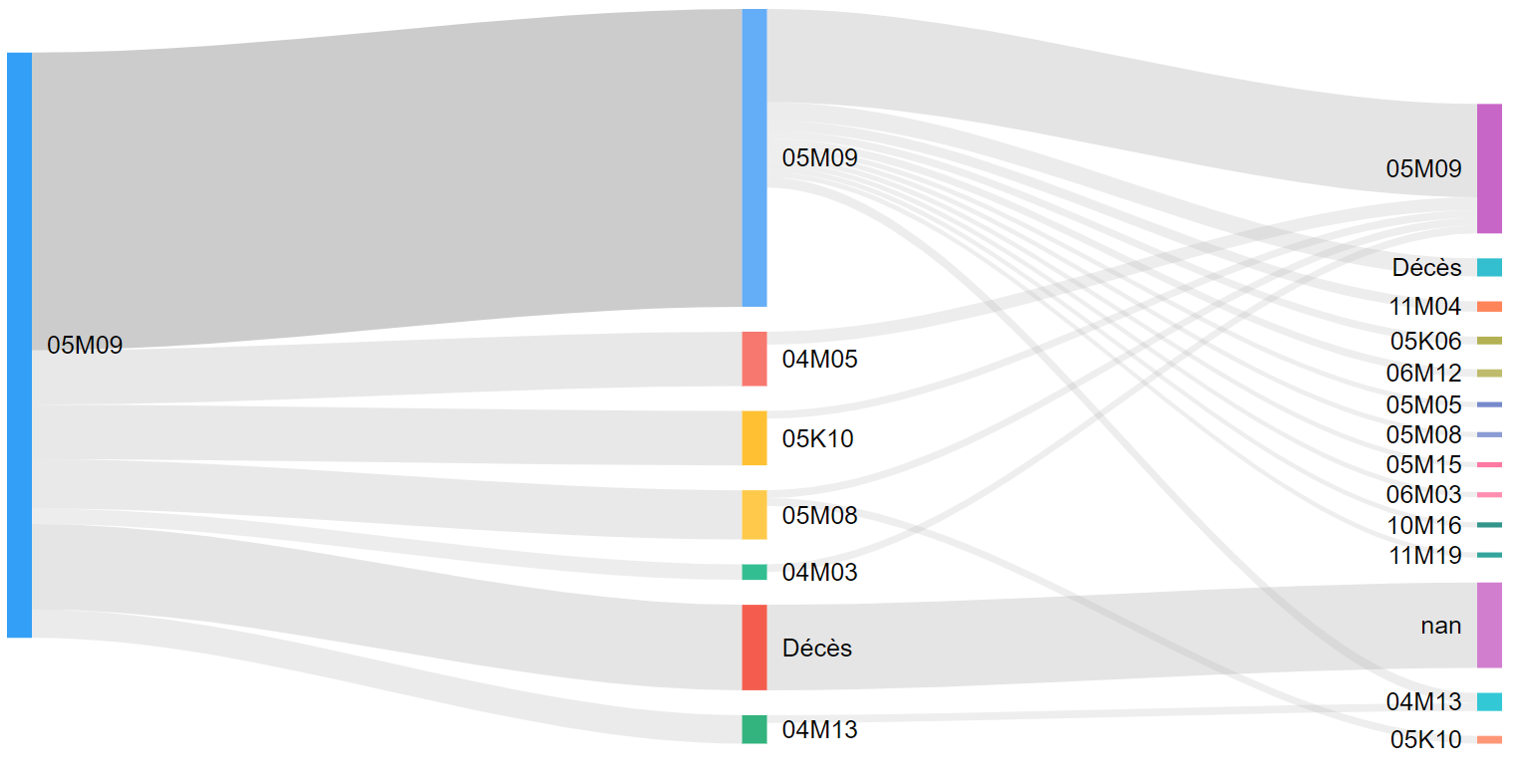}  
    \caption{Sankey Diagram Cluster 1}
\end{subfigure}
\begin{subfigure}{.5\textwidth}
    \centering
    \includegraphics[width=\linewidth]{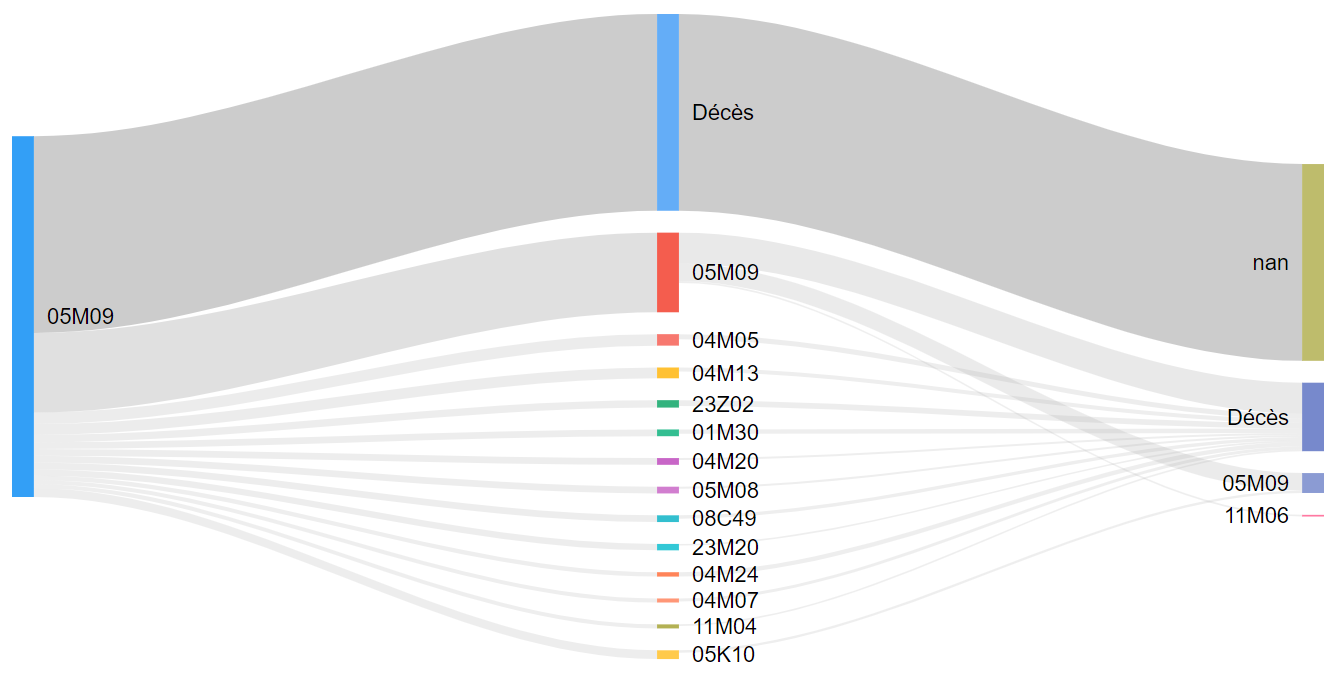}  
    \caption{Sankey Diagram Cluster 2}
\end{subfigure}
\begin{subfigure}{.49\textwidth}
    \centering
    \includegraphics[width=\linewidth]{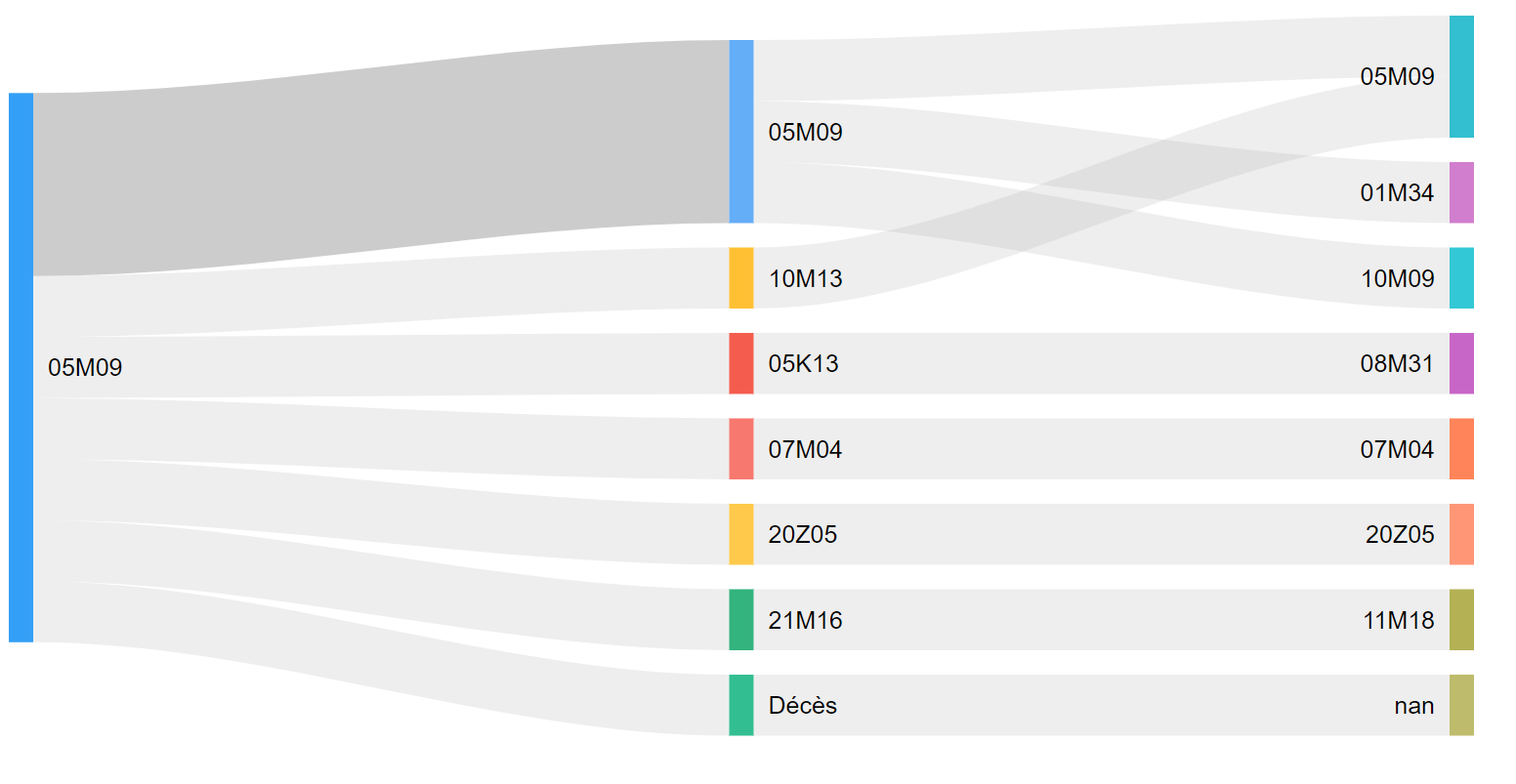}  
    \caption{Sankey Diagram Cluster 3}
\end{subfigure}
\begin{subfigure}{.49\textwidth}
    \centering
    \includegraphics[width=\linewidth]{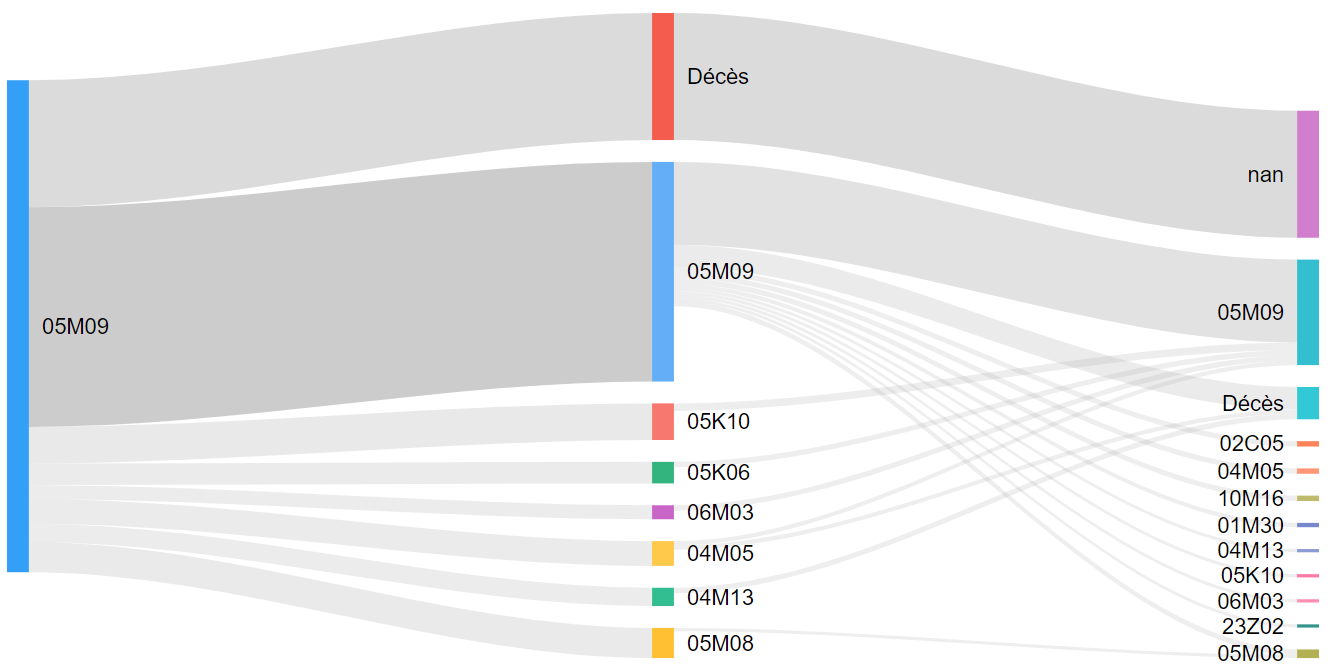}  
    \caption{Sankey Diagram Cluster 4}
\end{subfigure}
\begin{subfigure}{.49\textwidth}
    \centering
    \includegraphics[width=\linewidth]{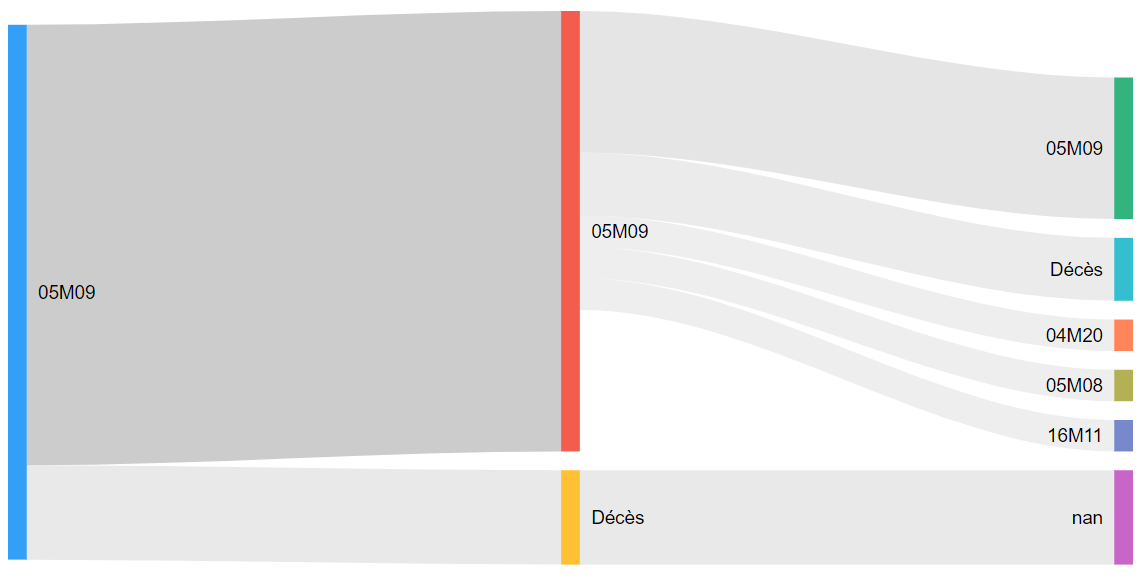}  
    \caption{Sankey Diagram Cluster 5}
\end{subfigure}
\caption{Sankey Diagrams: patient frequent flows at the cluster level}
\end{figure}

\newpage
Here is the Python code for the \texttt{generateSankey} function using the \texttt{prefixspan} library:
\begin{lstlisting}[caption={Python code for generating Sankey diagrams using PrefixSpan}, label=python_code]
from prefixspan import PrefixSpan

def generateSankey(df_care_pathways, cluster_n, top_k_frequent_codes):
    # df_care_pathways represents the pathways of deceased heart-failure patients
    df_pathways_cluster = df_care_pathways.loc[df_care_pathways.cluster == cluster_n]
    nb_patients_cluster_n = len(df_pathways_cluster)
    df_pathways_23 = df_pathways_cluster.astype(str).iloc[:,2:4] 
    # keep first and second ICD-10 codes
    df_pathways_34 = df_pathways_cluster.astype(str).iloc[:,3:5]
    # keep second and third ICD-10 codes
    
    corpus_heart_failures = []
    for patient in range(nb_patients_cluster_n):
        icd_codes = [df_pathways_23.iloc[patient][0], df_pathways_23.iloc[patient][1]]
        corpus_heart_failures.append(icd_codes)

    ps = PrefixSpan(corpus_heart_failures)
    ps.minlen = 2
    output_corpus = ps.topk(top_k_frequent_codes)

    corpus_heart_failures_2 = []
    for patient in range(nb_patients_cluster_n):
        if str(df_pathways_34.iloc[patient][1])!='nan':
            icd_codes = [df_pathways_34.iloc[patient][0], df_pathways_34.iloc[patient][1]]
            corpus_heart_failures_2.append(icd_codes)
        else:
            corpus_heart_failures_2.append([df_pathways_34.iloc[patient][0], 'nan'])

    ps_corpus_2 = PrefixSpan(corpus_heart_failures_2)
    ps_corpus_2.minlen = 2
    output_corpus_2 = ps_corpus_2.topk(top_k_frequent_codes)

    return output_corpus, output_corpus_2
\end{lstlisting}

The outputs of the function \texttt{generateSankey} above are then used with HTML code to create the clean Sankey outputs.

\begin{figure}[h]
    \centering
    \includegraphics[width=8.4cm]{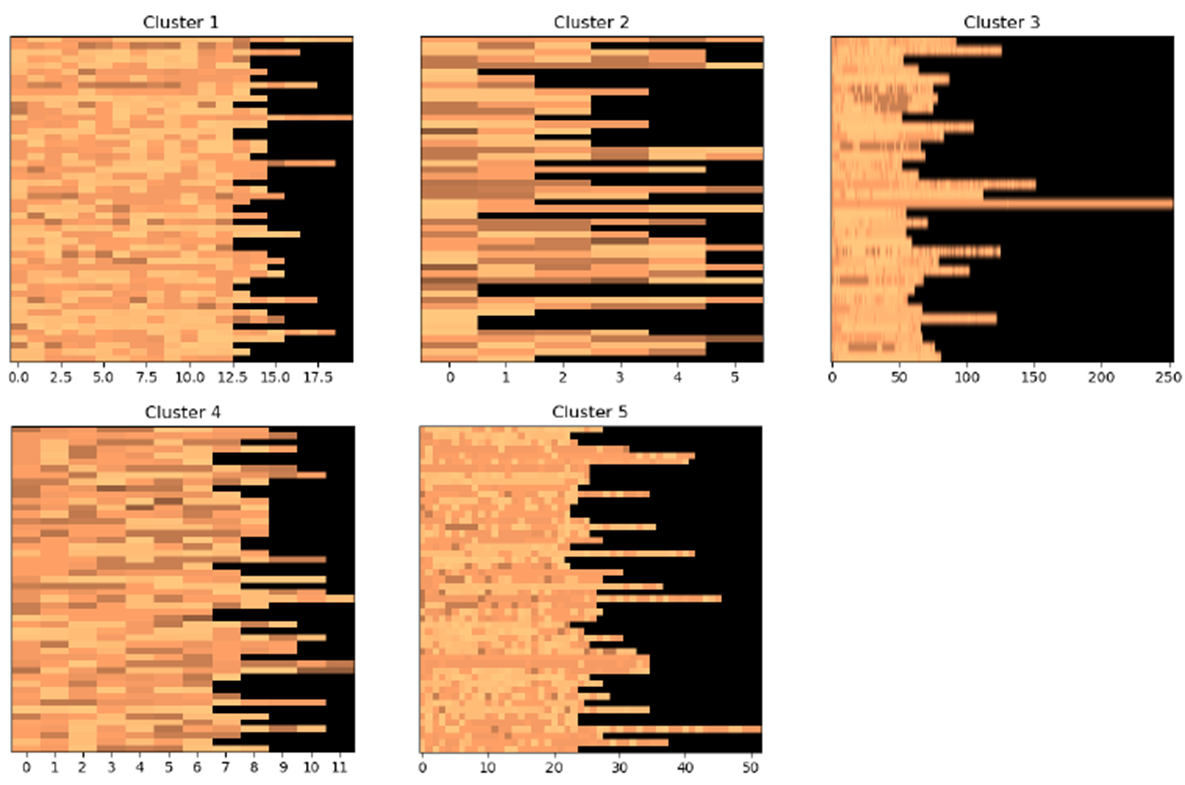}
    \caption{Distance of ICD-10 codes with the medoid for each cluster}
    \label{fig:clust2}
\end{figure}

We also evaluated the clusters by calculating the minimum distance between the ICD-10 code of each data point and the ICD-10 codes of the cluster's medoid (see Figure \ref{fig:clust2}). The shading in the graphs indicates the extent to which the ICD-10 code deviates from the hospitalization pattern of the patient medoid. Lighter areas, particularly in clusters 1 and 5, suggest that the data points within these clusters exhibit relatively close similarity to one another.

\subsection{Survival analysis}
We define the survival function 
\begin{equation*}
    S(t) = \mathbb{P}[T > t] = 1 - F(t) = \int_t^\infty f(u)du
\end{equation*}
with $f$ the density and $F$ the distribution function. $T$ is the time of occurrence of some specific event, time to death in our case. We introduce censoring $C$ to the labels, whenever the patient died or not with $T^* = \min (T, C)$ and $\Delta = \mathbb{1}_{\{T \leq C\}}$. 

A convenient way of modeling relationships between $T$ and explanatory features $X$ is the hazard function. Let $\lambda(t|X)$ denote the hazard function associated with the distribution of $T$ given $X$. 

The Cox proportional hazard (CPH) is the most common approach to model survival events \citep{cox1972regression}. It writes
\begin{equation*}
    \lambda(t|X) = \lambda_0^*(t)\exp(f^*(X))
\end{equation*}
$\lambda_0^*$ is the baseline hazard function and usually $f^*(X) = X^T\beta$ with $\beta \in \mathbb{R}^d$. 

This model assumes hazards to be proportional, i.e. hazard ratios to be independent of time as 
\begin{equation*}
\frac{\lambda_i(t|X_i = x_i)}{\lambda_j(t|X_j = x_j)} = \exp(f^*(x_i)-f^*(x_j))    
\end{equation*}

This approach can be penalized and/or include splines to handle non-linear relations \citep{tibshirani_regression_1996}.
This semi-parametric approach was applied to the data, however its main assumptions were never met. For this reason ensemble methods were selected.

Here is the Python code for the \texttt{Survival random forest} function using the \texttt{scikit survival} library:
\begin{lstlisting}[caption={Python code for generating the Grid Search over the number of estimator in our survival random forest}, label=python_code]
data=profil_patient[['y_nais','BEN_SEX_COD','Nb_hospit','CHOC','Nb_jours_sej']]
Label= profil_patient[['Mort','Nb_survie']].to_records(index=False)

X_train, X_test, y_train, y_test = train_test_split(data, Label, test_size=0.25)
Liste_score = []
Liste_nb_estimators= []
rsf = RandomSurvivalForest(min_samples_split=10,
                           min_samples_leaf=15,
                           n_jobs=-1,
                           random_state=random_state,
                           verbose=0)
for i in range(1, 20):
    n_estimators = i * 1
    rsf.set_params(n_estimators=n_estimators)
    rsf.fit(X_train, y_train)
    Liste_score.append(rsf.score(X_test, y_test))
    Liste_nb_estimators.append(n_estimators)
    print(rsf.score(X_test, y_test))
    

plt.plot(Liste_nb_estimators, Liste_score)
plt.xlabel("")
plt.ylabel("")
plt.grid(True)

\end{lstlisting}

\begin{table}[h]
\caption{Evaluation for each cluster} \label{tab:surv-clust}
\begin{center}
\begin{tabular}{l|ll}
\textbf{Cluster}  & \textbf{AIC} & \textbf{C-Index}\\
\hline 
1 & 8,495    & 0.612 \\
2 & 41,748   & 0.640 \\
3 & 39       & 0.887 \\
4 & 24,015   & 0.621 \\
5 & 1773     & 0.543\\
\end{tabular}
\end{center}
\end{table} 

\subsection{Clinical discussion}\label{app:cd}
The cohort of this study was comparable to other similar studies, namely in terms of age and gender, and with poor prognosis with high rate of hospitalization and mortality \citep{farre_real_2017}. The results obtained from this work may thus be comparable to those of other similar groups. For this reason, age and being male being strongly associated with mortality were expected findings \citep{simpson2020prognostic}.

Nevertheless, our study is subject to several limitations. First, the inclusion of primary diagnoses alongside the principal diagnoses in hospitalizations could provide a more comprehensive view of patients' medical histories, enabling a deeper understanding of their healthcare trajectories. Unfortunately, acquiring such data presents a challenge, as only a limited subset of hospitalizations in our dataset contains this information. Consequently, our analysis is constrained to ICD-10 codes. Besides, the sequential pattern mining analysis uncovered certain high-risk patterns that could potentially serve as additional features in our survival model, similar to our inclusion of cardiac shock. Integrating these factors into our model would enable to quantify the excess mortality associated with these risky healthcare sequences. We believe that this additional information could enhance the precision of our predictions and yield a more nuanced understanding of the underlying causes of mortality among heart failure patients.

\end{document}